\documentclass[12pt, a4paper, reqno]{amsart}
\usepackage{amssymb, amsmath, mathrsfs, amsthm}
\usepackage{graphicx}
\usepackage{color}
\usepackage[top=1in, bottom=1in, left=1in, right=1in]{geometry}
\usepackage{float, caption, subcaption}
\usepackage{graphicx}%
\usepackage{multirow}%
\usepackage{amsmath,amssymb,amsfonts}%
\usepackage{amsthm}%
\usepackage[title]{appendix}%
\usepackage{xcolor}%
\usepackage{textcomp}%
\usepackage{manyfoot}%
\usepackage{booktabs}%
\usepackage{algorithm}%
\usepackage{algorithmicx}%
\usepackage{algpseudocode}%
\usepackage{listings}%
\usepackage{float}%
\usepackage{enumitem}%
\usepackage{url}%
\usepackage{hyperref}

\DeclareMathOperator{\ESh}{ESh}
\DeclareMathOperator{\Sh}{Sh}

\theoremstyle{plain}%
\newtheorem{theorem}{Theorem}%
\newtheorem{proposition}[theorem]{Proposition}%
\newtheorem{lemma}[theorem]{Lemma}%
\newtheorem{corollary}[theorem]{Corollary}%

\theoremstyle{plain}%
\newtheorem{definition}[theorem]{Definition}%
\newtheorem{example}[theorem]{Example}%

\theoremstyle{plain}%
\newtheorem{remark}[theorem]{Remark}%

\begin{document}
\title[]{An Edge-based Shapley value \\ for supply chain network analysis}

\author[T. Yamada]{Taiki Yamada$^\dag$}
\address{Interdisciplinary Faculty of Science and Engineering, Shimane University.}
\email[Corresponding author]{taiki\_yamada@riko.shimane-u.ac.jp}
\thanks{$\dag$ corresponding author}

\author[T. Matsubae]{Taisuke Matsubae}
\address{Faculty of Economics, Chuo University.}
\email{matsubae@gmail.com}

\author[T. Akamatsu]{Tomoya Akamatsu}
\address{MUFG Bank, Ltd.}
\email{t.akamatsu.05c65@gmail.com}

\pagestyle{plain}

\subjclass[2020]{Primary~68R10, Secondary~90B06}
\keywords{Shapley value, Supply chain analysis, Network science.}
\begin{abstract}
This study introduces an \emph{edge-based Shapley value}, a novel allocation rule in cooperative game theory tailored specifically to supply chain networks, where value is generated through edge-mediated interactions.
Traditional allocation rules, such as the Shapley value and Myerson value, evaluate player contributions based on node-level characteristics or connected components.
However, these approaches often fail to adequately capture the functional role of edges that represent supply routes with associated costs and flow volumes.
Our edge-based Shapley value shifts the characteristic function from node sets to edge sets, thereby enabling a more granular and context-sensitive evaluation of supplier contributions.
We establish its theoretical foundations, demonstrate its relationship to classical allocation rules, and show that it retains key properties such as fairness and symmetry.
We apply the method to supply chain networks by incorporating route-specific supply quantities and transportation costs via a cost-decaying weight function, and validate the approach through a systematic empirical benchmark on \emph{seven distinct supply network topologies}, spanning serial, parallel-redundant, asymmetric tier, scale-free, clustered, layered DAG, and single-point-of-failure structures.
In simultaneous multi-node failure experiments, ESV-guided removal of the top-$3$ nodes destroys an average of $93.4\%$ of the total supply value---substantially outperforming betweenness centrality ($84.5\%$) and degree centrality ($88.1\%$).
Furthermore, we show that ESV rankings are more robust to prior network disruption than the single-node removal measure $\Delta_i$: when portions of the network have already failed, the ESV computed on the original intact network predicts the remaining nodes' importance more accurately than $\Delta_i$, because the Shapley value inherently averages over all possible degradation states.
A comparison with the efficiency-based vulnerability measure of Latora and Marchiori~\cite{latora2001} reveals that the ESV and betweenness centrality are \emph{complementary}: the former captures economic supply-value loss incorporating route volumes and costs, while the latter captures topological connectivity loss.
This framework provides practical tools for vulnerability assessment and value attribution in supply chain networks.

\medskip
\noindent\textbf{Keywords:} Shapley value; Supply chain analysis; Network science; Game theory; Graph theory; Vulnerability assessment.
\end{abstract}
\maketitle

\section{Introduction}
\subsection{Background}
Allocation rules within cooperative game theory have consistently provided a robust framework for understanding the equitable distribution of value among agents engaged in collaborative activities.
Notably, the Shapley value, introduced by Shapley \cite{Sh}, is well known for its axiomatic foundation.
It assigns to each player an expected marginal contribution, averaged across all potential coalition formation sequences, and adheres to essential properties such as efficiency, symmetry, and the null player condition.
Consequently, the Shapley value has become a widely used tool in disciplines ranging from economics and operations research to machine learning and explainable AI \cite{lundberg2017unified, strumbelj2014explaining}. 
 
However, many real-world applications involve networks in which interactions between agents are not arbitrary but are constrained by an underlying graph structure.
For instance, in supply chains and transportation networks, agents interact along specific edges that represent contracts, material flow, or logistical links.
To incorporate such structural constraints into cooperative games, Myerson \cite{My} proposed a modification of the Shapley value, now known as the Myerson value, which limits feasible coalitions to connected components of the graph.
Although this was a significant advancement, the Myerson value still evaluates coalitions at the node level and relies heavily on component-based aggregation, which may obscure critical information encoded in edge-level interactions. 
 
Recent developments in network science and algorithmic game theory emphasize the importance of modeling value generation not merely through node-level participation but through the structural roles of edges that connect agents and facilitate their interaction.
In particular, edge-centric models have proven essential in systems where value arises from flows of goods, information, or influence rather than static presence.
For example, in supply chain analysis, the significance of a supplier often derives not from its centrality in the network, but from the specific supply routes it enables, as well as the costs and risks associated with those paths \cite{bell2021resilience, tang2006perspectives}. 
Recent methodological advances in edge attribution and graph-based influence estimation further reinforce the view that edge-level modeling yields richer and more actionable insights than traditional node-centric approaches \cite{agarwal2022edge, rossi2020proximity}. 

These perspectives motivate the need for cooperative game-theoretic tools that operate directly on edge structures, thereby enabling more granular and realistic evaluations of agent contributions in networked systems.
To address these shortcomings, we propose a new allocation rule, the \emph{edge-based Shapley value}, which shifts the characteristic function from node sets to edge sets.
This allows for a more granular and context-sensitive assessment of contributions, particularly in systems where value is not generated by mere inclusion in a coalition but by participating in structured interactions.
Our approach captures route-level importance, accommodates non-additive value structures, and enables a deeper understanding of overlapping roles in complex interdependent networks.
 
\subsection{Contribution of this paper}
This study makes the following key contributions to the theory and application of cooperative games on networks:
\begin{itemize}[leftmargin=*]
  \item \textbf{Theoretical formulation.} We introduce the edge-based Shapley value,
a novel allocation rule that defines the characteristic function on edge subsets.
We show that this formulation generalizes the Myerson value and satisfies desirable axiomatic properties, such as symmetry, fairness, and a modified form of component efficiency.
  \item \textbf{Comparative analysis.} We provide a formal comparison between the
proposed edge-based Shapley value and classical allocation rules, including the Shapley value and Myerson value, and highlight the limitations of node-based methods in capturing edge-level interactions.
  \item \textbf{Application to supply chains.} We apply the edge-based Shapley value
to supply chain networks by incorporating route-specific supply quantities and transportation costs via a cost-decaying weight function.
  \item \textbf{Systematic empirical benchmark.} We design a benchmark of seven qualitatively distinct supply network topologies and evaluate the ESV through three complementary analyses: (i) simultaneous multi-node failure experiments, in which ESV identifies the most disruptive node sets more effectively than betweenness and degree centrality; (ii) a predictive robustness test, in which ESV computed on the intact network predicts node importance in degraded networks more accurately than the single-node removal measure $\Delta_i$; and (iii) a comparison with the Latora--Marchiori efficiency-based vulnerability $V_i^{\rm eff}$, which reveals that ESV and betweenness centrality are complementary.
\end{itemize}
 
\section{Preliminaries}\label{sec:prelim}
This section defines the basic concepts necessary for this study.
\subsection{Graph Theory}
A graph $G = (N, E)$ is represented by a pair consisting of a set of players $N = \left\{1, \ldots, n \right\}$ and a set of edges $E$. An \textit{edge} $e \in E$ is an ordered pair $e = (u, v)$, where $u, v \in N$. 
 
\begin{definition}
Let $G=(N, E)$ be a graph. 
For any player $u \in N$, the \textit{neighborhood} of $u$ is defined by
\begin{eqnarray*}
\Gamma^G (u) = \left\{ v \in N \mid (u, v) \in E \right\}.
\end{eqnarray*}
\end{definition}
The neighborhood of a player $i$ refers to the set of players adjacent to $i$. 
It is highly likely that players who are directly related to a player will form a coalition, but it is also possible that players who are indirectly related to the player will form a coalition through the mediation of a certain player. 
Thus, we define a path as follows:
\begin{definition}
For any two players $s$ and $t$, a \textit{path} from $s$ to $t$ is a sequence of edges $\left\{ (v_i, v_{i+1}) \right\}_{i=0}^{n-1}$, where $v_0 = s,\ v_n = t$.
\end{definition}
\begin{remark}
If there exists a path between $s$ and $t$, then $s$ and $t$ are called \textit{connected}. 
In particular, if any pair of players in $N$ is connected, then $G=(N, E)$ is called \textit{connected}. 
\end{remark}

\begin{definition}
Let $G = (N, E)$ be a graph. For a subset $S \subseteq N$, the \textit{induced subgraph} $G[S]$  is the pair of $(S, E_S)$, where $E_S$ satisfies $E_S = \left\{ e \in E \mid e \subseteq S \right\}$.
\end{definition}
A maximal connected induced subgraph is called a \textit{connected component} of $G$, and $(N/E)$ denotes the set of all connected components in $G$. Note that the union of all connected components in $G$ is equal to $N$. 
 
\subsection{Game Theory}
A TU game is a pair $(N, v)$, where $N$ denotes a set of players and $v: 2^N \to \mathbb{R}$ is a characteristic function with $v(\emptyset) = 0$. The set of all TU games is denoted by $TU(N)$. Each $S \in 2^N$ is referred to as a \textit{coalition}, and $v(S)$ represents the \textit{worth} of $S$.  
 
\begin{definition}
A \textit{graph game} is a triplet $(N, v, E)$, where $(N, v)$ is a TU game and $(N, E)$ is a graph. 
The set of all graph games is denoted by $GG(N)$.
\end{definition}
 
We now define the Shapley value, which is a well-known allocation rule.
	\begin{eqnarray*}
	Sh_i (N, v) = \sum_{S \subseteq N \setminus \left\{i \right\}} \cfrac{s! (n - s - 1)!}{n!} (v(S \cup \left\{i \right\}) - v(S)).
	\end{eqnarray*}
Although many properties are related to the Shapley value, the focus of this study is on the following characterization theorem.
\begin{theorem}[\cite{Sh}]
\label{thm:Shapley}
The Shapley value is the unique allocation rule that is efficient, additive, symmetric, and satisfies to the null player property, subject to the following conditions:
\begin{enumerate}
\item An allocation rule $f$ on $TU(N)$ is \textit{efficient} if for any $(N, v) \in TU(N)$,
	\begin{eqnarray*}
	\sum_{i \in N} f_i(N, v)= v(N).
	\end{eqnarray*}
\item An allocation rule $f$ on $TU(N)$ is \textit{additive} if for all $(N, v), (N, w) \in TU(N)$,
	\begin{eqnarray*}
	f_i (N, v + w) = f_i (N, v) + f_i (N, w)
	\end{eqnarray*}
	for any player $i \in N$.
\item An allocation rule $f$ on $TU(N)$ is \textit{symmetric} if for any $(N, v) \in TU(N)$,
	\begin{eqnarray*}
	f_i(N, v) = f_j (N, v)
	\end{eqnarray*}
	for any two players $i, j \in N$ with $v(S \cup \left\{ i \right\}) = v(S \cup \left\{j \right\})$ for all $S \subseteq N \setminus \left\{i, j \right\}$.
\item An allocation rule $f$ on $TU(N)$ satisfies the \textit{null-player property} if for any $(N, v) \in TU(N)$,
	\begin{eqnarray*}
	f_i (N, v) = 0
	\end{eqnarray*}
	for any $i \in N$ with $v(S) = v(S\cup \left\{i \right\})$ for all $S \subseteq N$.
\end{enumerate} 
\end{theorem}
 
The \textit{Myerson value} $\mu (N, v, E)$ for undirected graphs can be expressed as
	\begin{eqnarray*}
	\mu_i (N, v, E) = Sh_i (N, v^E) = \sum_{S \subseteq N \setminus \left\{i \right\}} \cfrac{s! (n - s - 1)!}{n!} (v^E(S \cup \left\{i \right\}) - v^E(S)),
	\end{eqnarray*}
where $v^E (S) = \sum_{T \in S/E_S} v(T)$ for any coalition $S$. 
The Myerson value has the following characterization.
\begin{theorem}[\cite{My}]
\label{thm:Myerson}
For any graph game $(N, v, E)$, the Myerson value is the unique allocation rule that satisfies the conditions of component efficiency and fairness, where
\begin{enumerate}
\item an allocation rule $f$ on graph games is \textit{component efficient} if for any $(N, v, E) \in GG(N)$,
	\begin{eqnarray*}
	\sum_{i \in T} f_i (N, v, E) = v(T)
	\end{eqnarray*}
	for any connected component $T \in N/E$, and
\item an allocation rule $f$ on the graph game is \textit{fair} if for any graph game $(N, v, E)$,
	\begin{eqnarray*}
	f_i (N, v, E) - f_i (N, v, E \setminus e) = f_j (N, v, E) - f_j (N, v, E \setminus e)
	\end{eqnarray*}
	for any $e \in E$ with $i, j \in e$.
\end{enumerate} 
\end{theorem}
 
\section{The edge-based Shapley value}\label{sec:esv}
\subsection{Definition}
\begin{definition}\label{def:egg}
An \emph{edge-based graph game} is a triplet $(N, E, w)$, where $(N, E)$ is a graph and $(E, w) \in TU(E)$.
The set of all such games is denoted by $EGG(N, E)$.
\end{definition}
 
For an edge-based graph game $(N, E, w)$, define the characteristic function $w^N \colon 2^{N} \to \mathbb{R}$ by
\begin{equation}\label{eq:wN}
w^N(S) = w\bigl(\{e \in E \mid e \subseteq S\}\bigr).
\end{equation}
\begin{remark}
The function $w^N$ satisfies the following properties.
\begin{enumerate}
\item For any player $u \in N$, $w^N(\left\{u \right\}) = w(\emptyset) = 0$.
\item $w^N(N) = w(E)$.
\item For any player $u \in N$, $w^N(N) - w^N(N \setminus \left\{u \right\})  = w(\left\{e \mid u \in e \right\})$.
\end{enumerate}
\end{remark}
 
\begin{definition}[Edge-based Shapley value]\label{def:esv}
For an edge-based graph game $(N, E, w)$, the edge-based Shapley value is
\begin{equation}\label{eq:esv}
\ESh_i(N, E, w) = \Sh_i(N, w^N) = \sum_{S \subseteq N \setminus \{i\}} \frac{s!\,(n-s-1)!}{n!}\bigl(w^N(S \cup \{i\}) - w^N(S)\bigr).
\end{equation}
\end{definition}
 
\begin{remark}\label{rem:esv-myerson}
For a graph game $(N, v, E)$, we define the characteristic function $\bar{w}: 2^{E} \to \mathbb{R}$ as follows.
\begin{eqnarray*}
    \bar{w}(\left\{e_1, \ldots, e_m \right\}) = v(\left\{s \mid s \in e_i\ \text{for some}\ i \right\}).
\end{eqnarray*}
Then it holds that for any subset $F \subseteq E$, 
\begin{eqnarray*}
    \bar{w}(F) = \sum_{i = 1}^k \bar{w}(F_c^i),
\end{eqnarray*}
where the disjoint union $\left\{F_c^1, \ldots, F_c^k \right\}$ satisfies the following conditions.
\begin{enumerate}
\item $F = F_c^1 \sqcup \cdots \sqcup F_c^k$.
\item For any $e \in F_c^i$, $e' \in F_c^j$, $e \cap e' = \emptyset$.
\end{enumerate}
In particular, for any coalition $S$, we have
\begin{eqnarray*}
\bar{w}^N (S)= v^E (S), 
\end{eqnarray*}
which implies
\begin{eqnarray*}
    ESh_i(N, E, \bar{w}) = \mu_i(N, v, E).
\end{eqnarray*}
The Myerson value is an allocation rule that is based on a characteristic function that depends only the connected components of the graph and is a special case of the edge-based Shapley value. 
In contrast, the edge-based Shapley value is an allocation rule that uses a characteristic function defined on the edges of the graph and can thus represent more complex and realistic situations.
\subsection{Comparison with Classical Allocation Rules}
Traditional allocation rules such as the Shapley and Myerson values evaluate player contributions based solely on node-level characteristics or connectivity, without fully capturing the role of intermediate structures within the network.
In contrast, the proposed edge-based Shapley value considers the structural significance of each participant by modeling supply routes as sets of edges and quantifying their marginal contributions accordingly.
This edge-centric approach facilitates a more precise evaluation of agent influence in scenarios in which the flow of goods, information, or services is inherently path-dependent, such as in supply chains. Furthermore, unlike the Myerson value, which necessitates partitioning the network into components and often assumes additive valuations, the edge-based Shapley value naturally accommodates overlapping and non-additive route values. This makes it particularly suitable for evaluating agents in layered and interdependent networks, where traditional rules may fail to differentiate between structurally symmetric but functionally distinct nodes. The simulation results in Section~\ref{sec:supply} further demonstrate that the edge-based Shapley value yields intuitive and consistent allocations that align well with the actual structural engagement of each agent within the supply network. Consequently, this method offers a refined and operationally meaningful alternative to classical allocation rules in networked cooperative settings, opening new avenues for both theoretical exploration and practical application in networked economic systems.
\end{remark}
\begin{figure}[ht]
     \centering
     \includegraphics[scale=0.45]{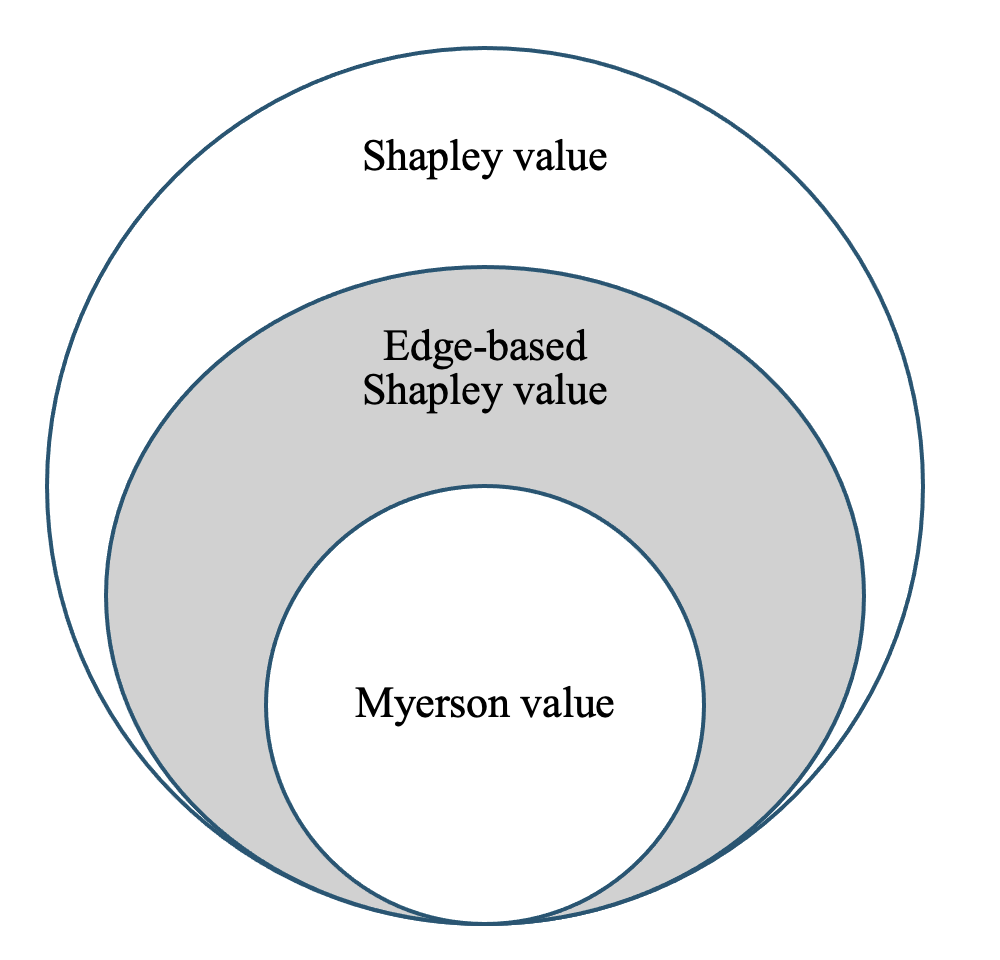} 
     \caption{Positioning of the edge-based Shapley value.}
     \label{position}
 \end{figure}

\section{Characteristics of the edge-based Shapley value}\label{sec:chars}
\subsection{Component efficiency}
\begin{proposition}
Suppose that the characteristic function $w^N$ satisfies
\begin{eqnarray*}
w^N (S \cup T) = w^N (S) + w^N(T)
\end{eqnarray*}
for any $S, T \subset N$ with $S \cap T = \emptyset$. Then we have
\begin{eqnarray*}
\sum_{i \in C}ESh_i (N, E, w) = w^N(C),
\end{eqnarray*}
for any connected component $C \in N/E$, and
\end{proposition}
 
\begin{proof}
Fix a connected component $C \in N/E$. We then consider the following characteristic function:
\begin{eqnarray*}
w^{C} (S) = \sum_{T \in (C \cap S)/ E_{C \cap S}} w^N (T)
\end{eqnarray*}
We remark that it holds that
\begin{eqnarray*}
w^{C} (C') = 0,
\end{eqnarray*}
for any $C \neq C' \in N/E$.
 
By this assumption, we have $w^N (S) = \sum_{C \in N/E} w^{C}(S)$, so we calculate as follows.
\begin{eqnarray*}
\sum_{i \in C} ESh_i (N, E, w) = \sum_{i \in C} \sum_{C' \in N/E} Sh_i (N, w^{C'}) = w^{C} (N) = w^N(C).
\end{eqnarray*}
\end{proof}
 
\begin{figure}[ht]
     \centering
     \includegraphics[scale=0.35]{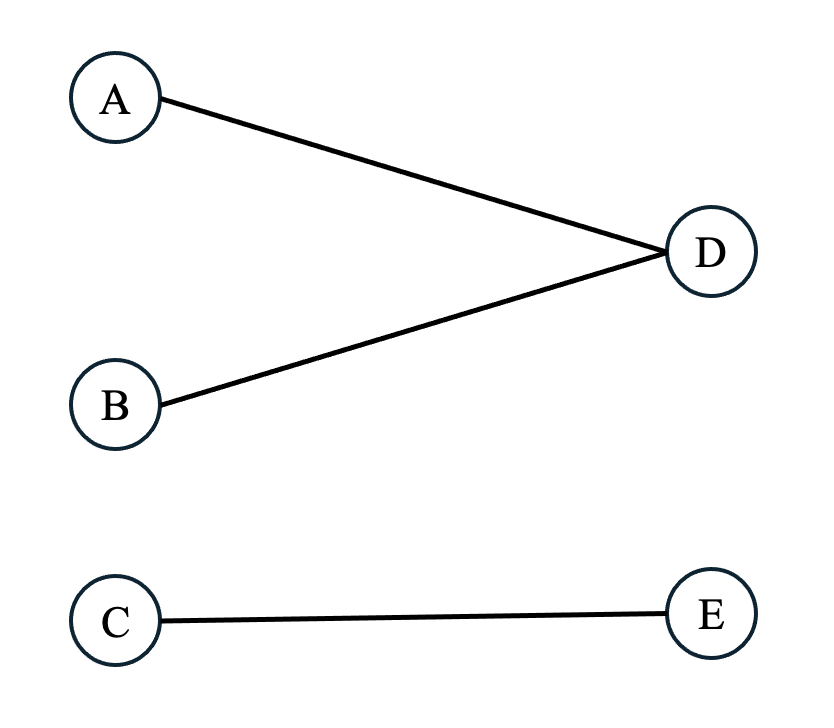} 
     \caption{Graph where Component efficiency does not hold.}
     \label{ex1}
 \end{figure}

\begin{example}
Considering the graph $H=(N, E)$ shown in Figure \ref{ex1}.
Then, it can be stated that
\begin{eqnarray*}
N &=& \left\{A, B, C, D, E \right\},\\
E &=& \left\{(A, D), (B, D), (C, E) \right\}.
\end{eqnarray*}
\end{example}
 
If the characteristic function is defined by
\begin{eqnarray*}
w(\left\{e_1, \ldots, e_n \right\}) = n^2,
\end{eqnarray*}
then the edge-based Shapley value is calculated as
\begin{eqnarray*}
ESh(N, E, w) = \left(\cfrac{5}{3}, \cfrac{5}{3}, \cfrac{3}{2}, \cfrac{8}{3}, \cfrac{3}{2} \right).
\end{eqnarray*}
Thus, we have
\begin{eqnarray*}
ESh_C (N, E, w) + ESh_E (N, E, w) = 3.
\end{eqnarray*}
However, $w^N(\left\{C, E \right\})= 1$.
 
\subsection{Fairness}
\begin{proposition}
For any edge-based graph game $(N, E, w)$, we have
	\begin{eqnarray*}
	ESh_i (N, E, w) - ESh_i (N, E \setminus e, w_e) = ESh_j (N, E, w) - ESh_j (N, E \setminus e, w_e)
	\end{eqnarray*}
	for any $e \in E$ with $i, j \in e$, and $w_e$ is defined by
    \begin{eqnarray*}
    w_e (S) = 
    \begin{cases}
    w(S),& \textrm{if}\ e \in S,\\
    w(S \setminus \left\{e \right\}),& \textrm{if}\ e \notin S.
    \end{cases}
    \end{eqnarray*}
\end{proposition}
\begin{proof}
For any edge $e = (i, j) \in E$, we set $z = w^N - w^N_e$. 
Then if $i \notin S$ or $j \notin S$, then we have
\begin{eqnarray*}
z(S) = w^N(S) - w^N_e (S) = w(\left\{e \mid e \subseteq S \right\}) - w_e(\left\{e \mid e \subseteq S \right\}) = 0,
\end{eqnarray*}
which implies
\begin{eqnarray*}
z(S \cup \left\{i \right\}) = z(S \cup \left\{j \right\}) =0
\end{eqnarray*}
for any $S \subseteq N \setminus \left\{i, j \right\}$. 
Thus, by the symmetry axiom of the Shapley value, it holds that $Sh_i (N, z) = Sh_j (N, z)$. 
By the additivity axiom of the Shapley value, it holds that
\begin{eqnarray*}
    ESh_i (N, E, w) - ESh_i (N, E \setminus e, w_e) &=& Sh_i (N, w^N) - Sh_i (N, w^N_e)\\
    &=& Sh_i (N, z) = Sh_j (N, z)\\
    &=& Sh_j (N, w^N) - Sh_j (N, w^N_e)\\
    &=& ESh_j (N, E, w) - ESh_j (N, E \setminus e, w_e).
\end{eqnarray*}
\end{proof}
 
\subsection{Original characteristics}
\begin{lemma}\label{lem:nbhd}
Fix a player $u \in N$. For any subset $S \subseteq N \setminus \Gamma^G (u)$,
\begin{eqnarray*}
w^N (S \cup \left\{u \right\}) - w^N (S) = 0
\end{eqnarray*}
\end{lemma}
\begin{proof}
Fix a subset $S \subseteq N \setminus \Gamma^G (u)$. For any player $v \in S$, we have
\begin{eqnarray*}
(u, v) \notin E,
\end{eqnarray*}
which implies
\begin{eqnarray*}
\left\{ e \mid e \subseteq S \right\} = \left\{ e \mid e \subseteq S \cup \left\{u \right\} \right\}.
\end{eqnarray*}
Thus, we obtain
\begin{eqnarray*}
w^N (S \cup \left\{u \right\}) = w^N(S).
\end{eqnarray*}
\end{proof}
 
Using this lemma, we obtain the following proposition.
\begin{proposition}\label{prop:reduced}
The edge-based Shapley value is represented as follows:
	\begin{eqnarray*}
    ESh_i (N, E, w) = \sum_{S \subseteq \Gamma^G (i)} \cfrac{s! (n - s - 1)!}{n!} (w^N (S \cup \left\{i \right\}) - w^N(S)).
   	\end{eqnarray*}
\end{proposition}
This reduced form, which restricts the sum to subsets of $\Gamma_G(i)$, will be crucial for the computational implementation of Section~\ref{sec:benchmark}.
 
\begin{corollary}
For any edge $e$ that contains a vertex $i$, we have
\begin{eqnarray*}
ESh_i (N, E \setminus e, w) = \sum_{S \subseteq \Gamma^G (i) \setminus \left\{j \right\}} \cfrac{s! (n - s - 1)!}{n!} (w^N(S \cup \left\{i \right\}) - w^N(S)),
\end{eqnarray*}
where $e = (i, j)$.
\end{corollary}
 
\section{Application to supply networks}\label{sec:supply}
\subsection{Definition}
\begin{definition}[Weighted supply network]
Let $G = (N, E, \ell)$ be a weighted graph, where $\ell \colon E \to \mathbb{R}_{\geq 0}$ assigns a non-negative transportation cost to each edge.
\end{definition}
 
\begin{definition}[Supply routes]
Let $\mathcal{R}$ be a set of supply routes in $G$.
For each route $r \in \mathcal{R}$, let $E(r) \subseteq E$ be the set of edges in $r$ and let $q_r \geq 0$ be the supply quantity transmitted along $r$.
The total transportation cost of $r$ is $c_r = \sum_{e \in E(r)} \ell(e).$
\end{definition}
 
\begin{definition}[Edge-set value function]
\label{def:walpha}
For a cost-sensitivity parameter $\alpha > 0$, define
\begin{equation}\label{eq:walpha}
w_\alpha(F) = \sum_{r \,:\, E(r) \subseteq F} q_r\,e^{-\alpha c_r}, \qquad F \subseteq E.
\end{equation}
\end{definition}
The exponential factor $e^{-\alpha c_r}$ models cost sensitivity: routes with larger total cost receive exponentially smaller weights, since $\partial e^{-\alpha c_r}/\partial c_r = -\alpha e^{-\alpha c_r} < 0.$ 
For small $\alpha$, supply quantities $q_r$ dominate; for large $\alpha$, low-cost paths are emphasized.
The exponential form ensures positivity, smooth monotonic decay, and preserves the relative ordering of routes, so $q_r e^{-\alpha c_r}$ can be interpreted as the \emph{effective supply contribution} of route $r$ after accounting for cost.
 
\subsection{Sensitivity and Robustness Analysis of the Edge-set value function}\label{sec:robustness}
To evaluate the robustness of the edge-based Shapley ranking, we consider two perturbation mechanisms:
\begin{enumerate}
    \item Parametric perturbation,
    $\alpha \in \{0.01,0.13,0.26,0.38,0.50\}$.
    \item Structural perturbation,
    random edge deletion with removal ratio\\
    $r \in \{0.0,0.1,0.2,0.3\}$.
\end{enumerate}
 
\begin{figure}[htbp]
\centering
\includegraphics[width=0.7\linewidth]{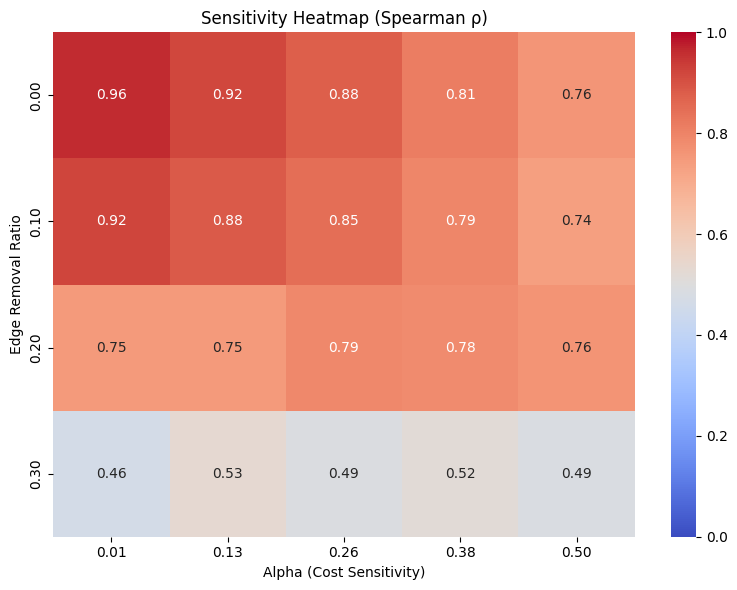}
\caption{
Spearman rank correlation $\rho(r,\alpha)$ between the baseline
edge-based Shapley ranking and the rankings obtained from perturbed networks.
Rows correspond to edge removal ratio $r$ and columns to
cost sensitivity parameter $\alpha$.
The ranking remains highly stable for $r \le 0.2$,
while substantial degradation occurs near $r=0.3$,
indicating the structural stability threshold.
}
\label{fig:sensitivity_heatmap}
\end{figure}

For each pair $(r, \alpha)$ we compute the Spearman rank correlation
\[
\rho(r, \alpha) = \rho_{\rm Spearman}\!\left(\ESh_{(r,\alpha)},\,\ESh_{(0,0.01)}\right),
\]
where $(0, 0.01)$ denotes the baseline network without edge removal and minimal cost sensitivity.
The results are summarized in the heatmap of Figure~\ref{fig:sensitivity_heatmap}.
 
The results satisfy $\min_\alpha \rho(0.0, \alpha) \geq 0.76$ and $\min_\alpha \rho(0.1, \alpha) \geq 0.74$, showing that the ranking is highly stable under mild structural perturbation.
For $r = 0.2$ we observe $\rho(0.2, \alpha) \approx 0.75$ (controlled variation), while at $r = 0.3$ the correlation drops to $\rho(0.3, \alpha) \in [0.46, 0.53]$.
Empirically, the correlation function satisfies $\partial\rho/\partial r < 0$, $\partial\rho/\partial \alpha < 0$, with $|\partial\rho/\partial r| > |\partial\rho/\partial \alpha|$.
The sharp decline near $r \approx 0.3$ suggests a structural threshold $r_c$ at which $\rho$ transitions from a high to a low regime, consistent with percolation-type behavior.
 
\section{Empirical validation on a diverse network benchmark}\label{sec:benchmark}
The sensitivity analysis in Section~\ref{sec:robustness} uses two stylized networks (a hub--spoke and a $3 \times 3$ grid).
Although these illustrate the qualitative behavior of the edge-based Shapley value, they cover only a small slice of the structural diversity encountered in real supply networks.
To demonstrate that the edge-based Shapley value (henceforth ESV) is broadly applicable to supply chain analysis, this section reports a systematic benchmark on \emph{seven qualitatively distinct supply network topologies} drawn from the operations research and network science literature.
\subsection{Benchmark networks}\label{sec:benchmark-nets}
The seven topologies are summarized in Table~\ref{tab:nets}.
Each is motivated by a well-known feature of real supply chains:
\begin{itemize}[leftmargin=*]
  \item \textbf{N1.\ Serial.} A single linear chain
$S \to v_1 \to \cdots \to T$, augmented with downstream sub-routes representing intermediate-product direct shipments.
This captures classical multi-echelon serial systems and the Beer Game family of models~\cite{tang2006}.
  \item \textbf{N2.\ Parallel redundant.} A source $S$ and a sink $T$ are
connected via $k$ disjoint two-edge paths through alternative middle suppliers, representing dual or multi-sourcing arrangements.
  \item \textbf{N3.\ Asymmetric tier.} A five-layer (suppliers $\to$ modules
$\to$ an assembler $\to$ distributors $\to$ retailers) chain with non-uniform fan-in and fan-out and a single integration point, mimicking automotive or consumer-electronics supply chains.
  \item \textbf{N4.\ Scale-free.} A Barab\'asi--Albert random graph
($n = 12$, $m = 2$), reproducing the heavy-tailed degree distribution that has been documented in inter-firm transaction networks.
  \item \textbf{N5.\ Clustered.} Three densely connected regional clusters
joined by a small number of expensive bridge edges, modeling geographically clustered supply with international logistics gateways.
  \item \textbf{N6.\ Layered DAG.} A randomly generated layered directed
acyclic graph with five layers of sizes $(3,4,3,2,3)$ and edges drawn independently between adjacent layers; a generic stochastic model of multi-echelon supply chains.
  \item \textbf{N7.\ Single point of failure (SPOF).} Two redundant
sub-networks joined by a single intermediate node $X$, an extreme test case for critical-node detection.
\end{itemize}
 
\begin{table}[htbp]
\centering
\caption{Benchmark networks: number of nodes $n$, number of edges $m$,
number of supply routes $|\mathcal{R}|$, and grand-coalition value $w^N(N)$ at $\alpha = 0.5$.}
\label{tab:nets}
\small
\begin{tabular}{lcccc}
\toprule
Network & $n$ & $m$ & $|\mathcal{R}|$ & $w^N(N)$ \\
\midrule
N1\ Serial               &  6 &  5 &  5 & 16.481 \\
N2\ Parallel redundant   &  6 &  8 &  4 &  6.867 \\
N3\ Asymmetric tier      & 13 & 15 &  6 &  9.484 \\
N4\ Scale-free (BA)      & 12 & 20 & 15 & 44.678 \\
N5\ Clustered            & 12 & 20 &  6 & 12.284 \\
N6\ Layered DAG          & 15 & 23 &  9 & 12.161 \\
N7\ SPOF                 &  7 &  8 &  4 &  9.843 \\
\bottomrule
\end{tabular}
\end{table}
 
\begin{figure}[htbp]
\centering
\includegraphics[width=\textwidth]{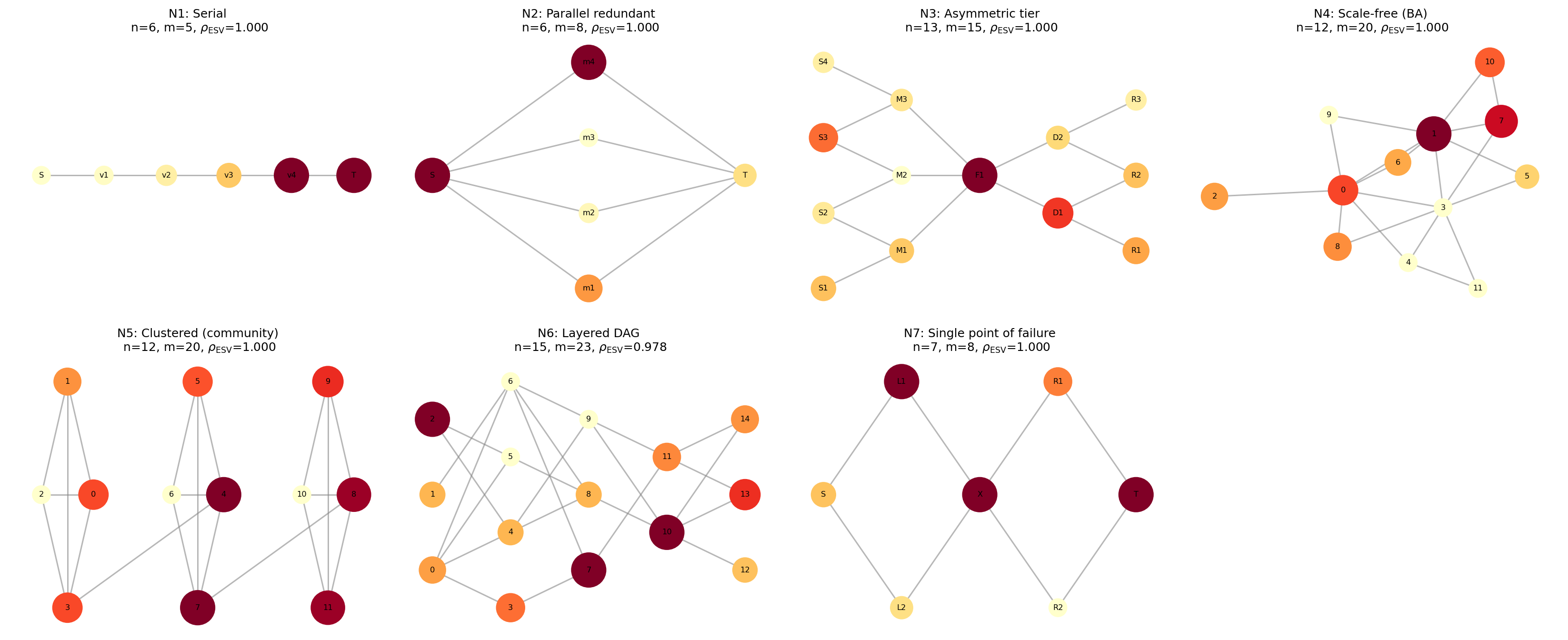}
\caption{The seven benchmark networks. The node color and size encode the edge-based Shapley value ($\ESh_i$), normalized within each panel.}
\label{fig:all-networks}
\end{figure}

For all networks, the characteristic function is the cost-sensitive supply function~\eqref{eq:walpha} of Section~\ref{sec:supply} with $\alpha = 0.5$.
Edge costs are drawn uniformly at random from $[0.5, 2]$ (with deterministic seeds for reproducibility), and the route set $\mathcal{R}$ is constructed explicitly for each topology so that all source--sink pairs are represented.
Figure~\ref{fig:all-networks} visualizes the seven networks; node color and size encode the ESV.
 
\subsection{Evaluation protocol}\label{sec:benchmark-eval}
We evaluate the ESV against three node-importance baselines:
\begin{enumerate}
    \item edge-weighted betweenness centrality, 
    \item degree centrality,
    \item a random ranking. 
\end{enumerate}
 
\subsection*{Evaluation measures.}
We evaluate node importance through two complementary analyses.
 
The first is a \emph{simultaneous multi-node failure} experiment.
For each importance score, we remove the top-$k$ nodes simultaneously and measure the residual supply value:
\begin{equation}\label{eq:residual}
R_k(\text{score}) \;=\; \frac{w^N\!\bigl(N \setminus \text{Top-}k(\text{score})\bigr)}{w^N(N)} \times 100\%.
\end{equation}
A lower $R_k$ indicates that the score has correctly identified the most disruptive node set.
This evaluation is operationally meaningful: real supply chain disruptions (earthquakes, pandemics, geopolitical events) often affect multiple nodes simultaneously, so the ability to identify the most damaging $k$-node failure set is directly relevant to risk management.
 
The second is a comparison with the \emph{efficiency-based vulnerability} of Latora and Marchiori~\cite{latora2001}:
\begin{equation}\label{eq:gt-external}
V_i^{\rm eff} = \frac{E_{\rm glob}(G) - E_{\rm glob}(G \setminus \{i\})}{E_{\rm glob}(G)}, \qquad E_{\rm glob}(G)
= \frac{1}{n(n-1)} \sum_{\substack{u,v \in N \\ u \neq v}} \frac{1}{d(u,v)},
\end{equation}
where $d(u,v)$ is the shortest-path distance in the edge-weighted graph.
This measure captures \emph{topological connectivity loss}: a node whose removal most degrades the average pairwise reachability is ranked highest.
Crucially, $V_i^{\rm eff}$ depends \emph{only} on graph topology and edge costs---it does not use the supply quantities~$q_r$, the route structure~$\mathcal{R}$, or the cost-sensitivity parameter~$\alpha$.
It is therefore \emph{mathematically independent} of the ESV.
This measure is widely used in infrastructure resilience and supply chain risk analysis~\cite{bocaletti2006,latora2003}.
 
For the comparison with $V_i^{\rm eff}$, we report Spearman's rank correlation $\rho$ between each importance score and $V_i^{\rm eff}$.
\subsection*{Exact computation.}
For all seven benchmark networks we compute the edge-based Shapley value \emph{exactly} according to Definition~\ref{def:esv}, exploiting the neighborhood-restricted summation of Proposition~\ref{prop:reduced}.
The largest network (N6, $n = 15$) requires at most $n \cdot 2^{n-1} \approx 2.46 \times 10^5$ evaluations of the characteristic function $w^N$, which finishes in well under one second on a standard laptop.
The total exact computation time across all seven networks is approximately $1.2$ seconds of wall-clock time; per-network timings are reported in Table~\ref{tab:timings}.
Because the evaluation is exact, the results below are \emph{reproducible deterministically} from the network seeds, and the ranking statistics are not subject to Monte Carlo sampling noise.
For reference, Appendix~\ref{app:mc} describes a permutation-based Monte Carlo estimator that can be used when exact computation becomes prohibitive ($n \gtrsim 25$).
 
\begin{table}[htbp]
\centering
\caption{Exact computation times for the edge-based Shapley value on the
seven benchmark networks, using the neighborhood-restricted formulation of Proposition~\ref{prop:reduced}.
Measurements were taken on a single CPU core of a standard laptop (Python 3.12, no parallelism).
The largest network, N6, has $n = 15$ nodes and takes less than $1$\,s.}
\label{tab:timings}
\small
\begin{tabular}{lccc}
\toprule
Network & $n$ & $n \cdot 2^{n-1}$ & Time (s) \\
\midrule
N1\ Serial             &  6 &        192 & $<0.001$ \\
N2\ Parallel redundant &  6 &        192 & $<0.001$ \\
N3\ Asymmetric tier    & 13 &     53{,}248 & 0.155 \\
N4\ Scale-free (BA)    & 12 &     24{,}576 & 0.075 \\
N5\ Clustered          & 12 &     24{,}576 & 0.190 \\
N6\ Layered DAG        & 15 &    245{,}760 & 0.791 \\
N7\ SPOF               &  7 &        448 & 0.001 \\
\midrule
Total                  & -- &         --   & $\approx 1.21$ \\
\bottomrule
\end{tabular}
\end{table}
 
\subsection{Results}\label{sec:benchmark-results}
 
\subsubsection{Simultaneous multi-node failure}
 
Table~\ref{tab:sim-removal} reports the residual supply value $R_k$ defined in~\eqref{eq:residual} when the top-$k$ nodes (according to each importance score) are removed simultaneously.
 
\begin{table}[htbp]
\centering
\caption{Residual supply value (\%) after simultaneously removing the top-$k$ nodes according to each importance score.
Lower values indicate better identification of critical node sets.
Bold entries indicate the best (lowest) score in each cell.}
\label{tab:sim-removal}
\small
\setlength{\tabcolsep}{4pt}
\begin{tabular}{l*{12}{c}}
\toprule
& \multicolumn{3}{c}{ESV (ours)}
& \multicolumn{3}{c}{Betweenness}
& \multicolumn{3}{c}{Degree}
& \multicolumn{3}{c}{Random} \\
\cmidrule(lr){2-4}\cmidrule(lr){5-7}\cmidrule(lr){8-10}\cmidrule(lr){11-13}
Network & $k{=}1$ & $k{=}2$ & $k{=}3$
        & $k{=}1$ & $k{=}2$ & $k{=}3$
        & $k{=}1$ & $k{=}2$ & $k{=}3$
        & $k{=}1$ & $k{=}2$ & $k{=}3$ \\
\midrule
N1 & \textbf{0.0} & \textbf{0.0} & \textbf{0.0} & 74.1 & 52.6 & 52.6 & 86.9 & 74.1 & 52.6 & \textbf{0.0} & \textbf{0.0} & \textbf{0.0} \\
N2 & \textbf{0.0} & \textbf{0.0} & \textbf{0.0} & \textbf{0.0} & \textbf{0.0} & \textbf{0.0} & \textbf{0.0} & \textbf{0.0} & \textbf{0.0} & \textbf{0.0} & \textbf{0.0} & \textbf{0.0} \\
N3 & \textbf{0.0} & \textbf{0.0} & \textbf{0.0} & \textbf{0.0} & \textbf{0.0} & \textbf{0.0} & \textbf{0.0} & \textbf{0.0} & \textbf{0.0} & 70.3 & 70.3 & 70.3 \\
N4 & \textbf{33.3} & 18.2 & \textbf{0.0} & 56.0 & \textbf{0.0} & \textbf{0.0} & 56.0 & \textbf{0.0} & \textbf{0.0} & 47.0 & 41.8 & 18.2 \\
N5 & \textbf{51.2} & \textbf{51.2} & \textbf{20.7} & \textbf{51.2} & \textbf{51.2} & 30.6 & 69.9 & 30.6 & 30.6 & 69.4 & 42.8 & 42.8 \\
N6 & \textbf{25.4} & \textbf{25.4} & \textbf{25.4} & \textbf{25.4} & \textbf{25.4} & \textbf{25.4} & \textbf{25.4} & \textbf{0.0} & \textbf{0.0} & 100.0 & 74.6 & 60.0 \\
N7 & \textbf{0.0} & \textbf{0.0} & \textbf{0.0} & \textbf{0.0} & \textbf{0.0} & \textbf{0.0} & \textbf{0.0} & \textbf{0.0} & \textbf{0.0} & 68.0 & \textbf{0.0} & \textbf{0.0} \\
\midrule
Mean & \textbf{15.7} & \textbf{13.5} & \textbf{6.6} & 29.5 & 18.5 & 15.5 & 34.0 & 14.9 & 11.9 & 50.7 & 32.8 & 27.3 \\
\bottomrule
\end{tabular}
\end{table}
 
At $k = 3$, ESV-guided removal destroys an average of $93.4\%$ of total supply value (residual $6.6\%$), compared with $84.5\%$ for betweenness (residual $15.5\%$) and $88.1\%$ for degree (residual $11.9\%$).
The advantage of ESV is most pronounced on networks where topological centrality fails to capture supply-flow structure: on N1 (serial), ESV achieves complete destruction at $k = 1$, while betweenness and degree leave $74\%$ and $87\%$ of supply intact, respectively.
On N5 (clustered), ESV at $k = 3$ reduces supply to $20.7\%$, whereas betweenness leaves $30.6\%$ and degree leaves $30.6\%$.
 
This result is non-trivial because the ESV is not designed to solve the combinatorial problem of finding the most disruptive $k$-node set; rather, it computes an axiomatic allocation of the total supply value $w^N(N)$ to individual nodes.
The fact that the ESV ranking nevertheless identifies highly disruptive multi-node failure sets demonstrates that the cooperative game-theoretic decomposition captures genuine structural dependencies among supply routes.
 
\subsubsection{Comparison with efficiency-based vulnerability \texorpdfstring{$V_i^{\rm eff}$}{V\textasciicircum eff\_i}}
 
\begin{figure}[htbp]
\centering
\includegraphics[width=\textwidth]{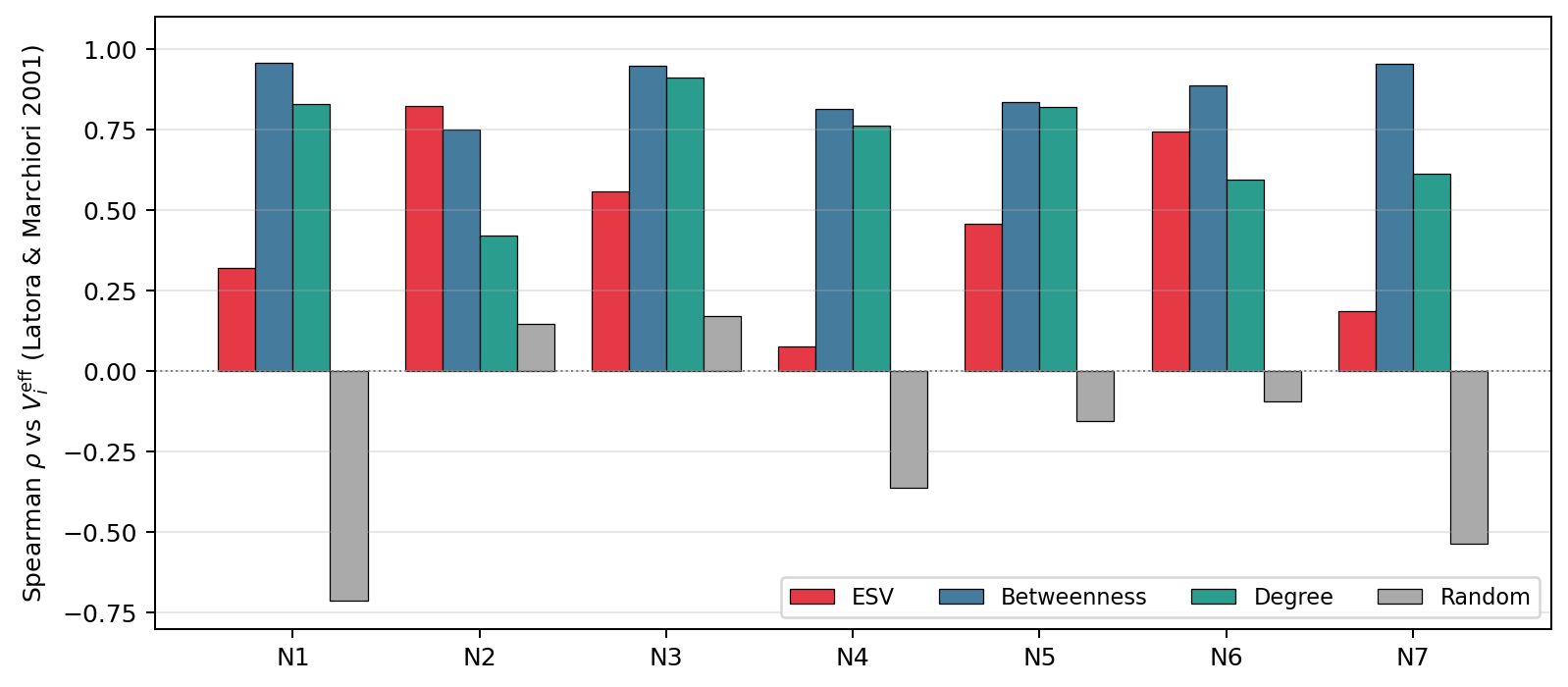}
\caption{Spearman rank correlation $\rho$ of each importance score with the efficiency-based vulnerability $V_i^{\rm eff}$ (external, independent of $w^N$).
Betweenness centrality dominates on six of the seven networks, revealing that the ESV captures a fundamentally different notion of node importance from topological centrality.}
\label{fig:dual-gt}
\end{figure}

Table~\ref{tab:spearman-ext} and Figure~\ref{fig:dual-gt} report the Spearman correlation of each score with the Latora--Marchiori efficiency-based vulnerability $V_i^{\rm eff}$ of~\eqref{eq:gt-external}.
 
\begin{table}[htbp]
\centering
\caption{Spearman rank correlation $\rho$ with the efficiency-based
vulnerability $V_i^{\rm eff}$ (external, independent of $w^N$).
Bold entries indicate the best score in each row.}
\label{tab:spearman-ext}
\small
\begin{tabular}{lcccc}
\toprule
Network & ESV (ours) & Betweenness & Degree & Random \\
\midrule
N1\ Serial             &  0.319 &  \textbf{0.956} &  0.828 & $-0.714$ \\
N2\ Parallel redundant &  \textbf{0.824} &  0.751 &  0.420 &  0.145 \\
N3\ Asymmetric tier    &  0.558 &  \textbf{0.948} &  0.912 &  0.170 \\
N4\ Scale-free (BA)    &  0.075 &  \textbf{0.815} &  0.762 & $-0.364$ \\
N5\ Clustered          &  0.457 &  \textbf{0.836} &  0.819 & $-0.154$ \\
N6\ Layered DAG        &  0.744 &  \textbf{0.887} &  0.595 & $-0.093$ \\
N7\ SPOF               &  0.185 &  \textbf{0.954} &  0.612 & $-0.536$ \\
\midrule
Mean                   &  0.452 &  \textbf{0.878} &  0.707 & $-0.221$ \\
\bottomrule
\end{tabular}
\end{table}
 
The picture is markedly different.
Betweenness centrality dominates on six of the seven networks (mean $\rho = 0.878$), whereas ESV achieves a mean of only $\rho = 0.452$.
The sole exception is N2 (parallel redundant), where ESV slightly outperforms betweenness ($0.824$ vs.\ $0.751$).
 
\subsection*{Supply degradation curves.}
\begin{figure}[htbp]
\centering
\includegraphics[width=\textwidth]{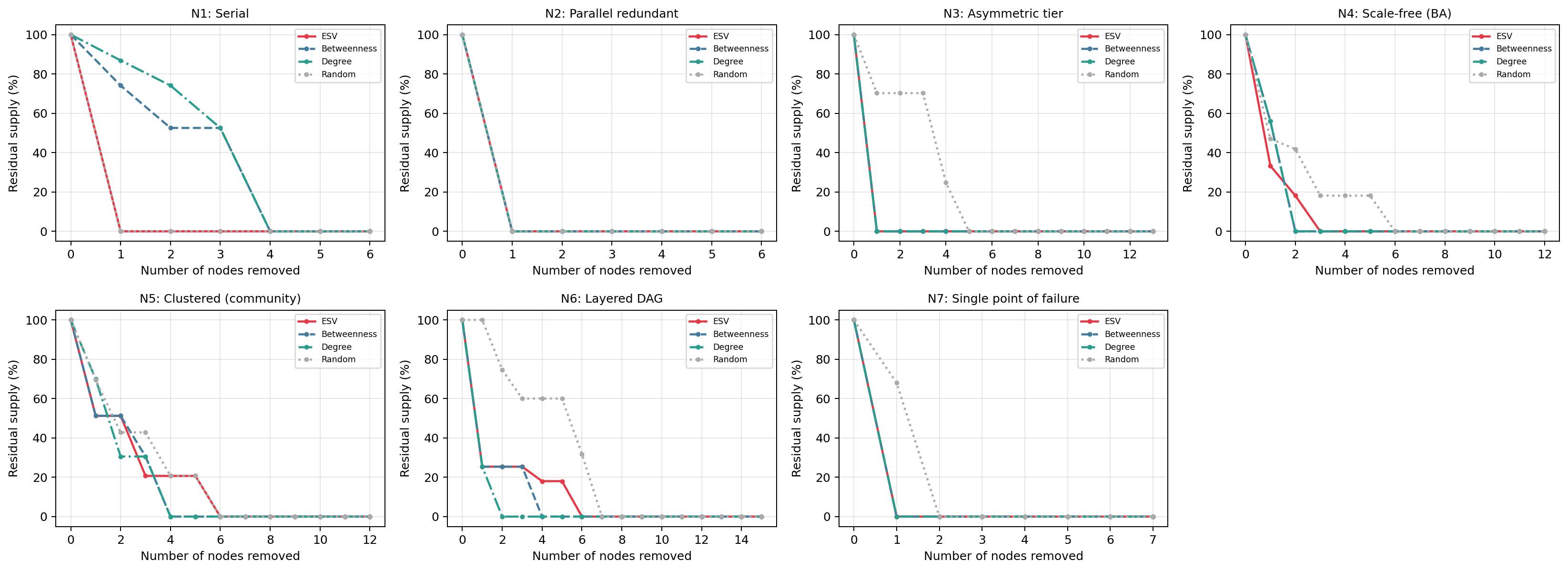}
\caption{Supply-value degradation under sequential node removal.
Lower curves indicate faster value destruction and hence better vulnerability identification.
The ESV consistently induces the steepest drop, confirming that it captures supply-route interdependencies that purely topological measures miss.}
\label{fig:degradation}
\end{figure}

Figure~\ref{fig:degradation} shows the residual supply value when nodes are removed one at a time in the order prescribed by each score.
The ESV curve is the steepest on every network, confirming that it identifies the most critical nodes earliest when the goal is to maximize the supply-value impact of removal.
 
\subsubsection{Predictive robustness under prior disruption}
 
To test whether the ESV provides more reliable pre-computed importance rankings than the single-node removal measure $\Delta_i = w^N(N) - w^N(N \setminus \{i\})$, we conduct the following experiment.
For each benchmark network, we (i) compute ESV, $\Delta_i$, and betweenness centrality on the \emph{intact} network; (ii) randomly remove $\lfloor r \cdot n \rfloor$ nodes to simulate prior disruption; (iii) on the degraded network, compute the ``true'' importance of each surviving node $j$ as $\Delta_j^{\rm deg} = w^N(S) - w^N(S \setminus \{j\})$ where $S$ is the set of survivors; and (iv) measure the Spearman rank correlation between each \emph{original} (pre-disruption) metric and the true degraded-network importance $\Delta_j^{\rm deg}$.
We repeat step (ii)--(iv) for $50$ random disruption samples and report the mean $\rho$.
A higher $\rho$ means the pre-computed metric remains a more reliable guide to prioritizing recovery efforts after unexpected disruption.
 
\begin{table}[htbp]
\centering
\caption{Predictive robustness: mean Spearman $\rho$ between pre-disruption importance scores and true importance on degraded networks, averaged over $50$ random disruption samples per network.
ESV achieves the highest $\rho$ on every network at every disruption rate, with the clearest advantage on the structurally complex N6 (Layered DAG).}
\label{tab:pred-robust}
\small
\begin{tabular}{llcccc}
\toprule
& & \multicolumn{4}{c}{Pre-disruption metric} \\
\cmidrule(lr){3-6}
Network & Rate $r$ & ESV (ours) & $\Delta_i$ & Betweenness & Degree \\
\midrule
\multirow{3}{*}{N1\ Serial}
  & 0.1 & \textbf{0.961} & 0.961 & $-0.222$ & $-0.176$ \\
  & 0.2 & \textbf{0.961} & 0.961 & $-0.222$ & $-0.176$ \\
  & 0.3 & \textbf{0.961} & 0.961 & $-0.222$ & $-0.176$ \\
\midrule
\multirow{3}{*}{N4\ Scale-free}
  & 0.1 & \textbf{0.918} & 0.918 &  0.097 & $-0.049$ \\
  & 0.2 & \textbf{0.860} & 0.860 &  0.083 & $-0.054$ \\
  & 0.3 & \textbf{0.793} & 0.793 &  0.069 & $-0.064$ \\
\midrule
\multirow{3}{*}{N6\ Layered DAG}
  & 0.1 & \textbf{0.593} & 0.581 &  0.450 &  0.077 \\
  & 0.2 & \textbf{0.547} & 0.531 &  0.401 &  0.072 \\
  & 0.3 & \textbf{0.475} & 0.468 &  0.375 &  0.061 \\
\midrule
\multirow{3}{*}{Mean (all 7)}
  & 0.1 & \textbf{0.821} & 0.819 &  0.342 &  0.290 \\
  & 0.2 & \textbf{0.752} & 0.750 &  0.317 &  0.273 \\
  & 0.3 & \textbf{0.689} & 0.688 &  0.323 &  0.251 \\
\bottomrule
\end{tabular}
\end{table}
 
\begin{figure}[htbp]
\centering
\includegraphics[width=\textwidth]{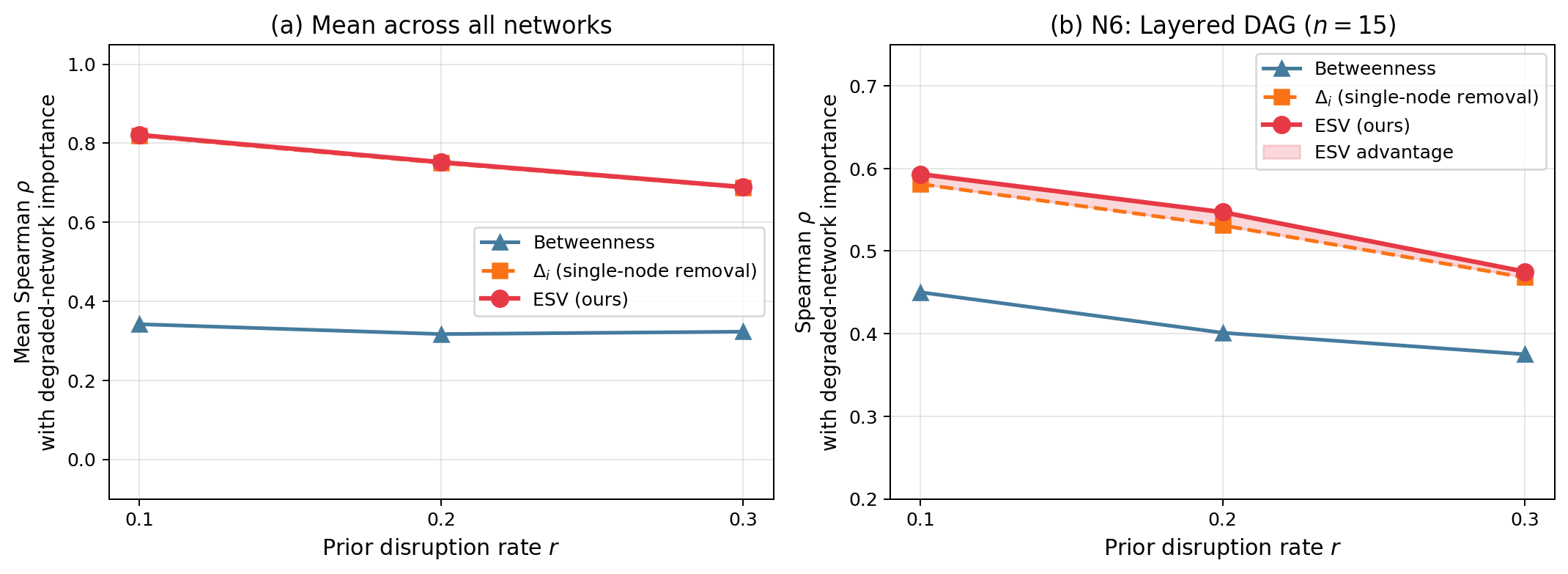}
\caption{Predictive robustness under prior disruption.
\textbf{(a)} Mean Spearman $\rho$ across all seven networks: ESV (red) consistently outperforms $\Delta_i$ (orange), with both substantially exceeding betweenness centrality (blue).
\textbf{(b)} Detail for the most structurally complex network N6 (Layered DAG, $n=15$): the shaded region highlights the ESV advantage over $\Delta_i$, which reaches $+0.016$ at $r = 0.2$.
The advantage of ESV arises because it averages marginal contributions across all coalition sizes, implicitly accounting for partially degraded network states.}
\label{fig:pred-robust}
\end{figure}
 
\subsection{Discussion}\label{sec:benchmark-discussion}
The benchmark results reveal three complementary findings: (i) the ESV excels at identifying economically disruptive multi-node failure sets; (ii) ESV rankings are more robust to prior network disruption than the single-node removal measure $\Delta_i$; and (iii) betweenness centrality better predicts topological connectivity loss.
 
\subsection*{Multi-node failure resilience.}
The simultaneous removal experiment (Table~\ref{tab:sim-removal}) demonstrates that ESV-guided node selection is the most effective at destroying supply value across the seven benchmark networks, achieving a mean residual of only $6.6\%$ at $k = 3$.
This result is noteworthy because the ESV is \emph{not designed} to solve the NP-hard problem of identifying the most disruptive $k$-node set.
Rather, it provides an axiomatic allocation of the total supply value $w^N(N)$ to individual nodes, satisfying the efficiency property $\sum_{i \in N} \ESh_i = w^N(N)$.
The strong performance in multi-node failure scenarios arises because the ESV accounts for route-level interdependencies---when two nodes share many supply routes, the ESV implicitly captures their joint criticality through the coalition structure of the Shapley value.
 
\subsection*{Predictive robustness under prior disruption.}
The predictive robustness experiment (Table~\ref{tab:pred-robust}; Figure~\ref{fig:pred-robust}) reveals a key empirical advantage of the ESV over the single-node removal measure $\Delta_i$.
Although both metrics yield similar rankings on the intact network, the ESV \emph{better predicts node importance in degraded networks}---i.e., networks that have already lost nodes to prior disruption.
On the most structurally complex network in the benchmark (N6, Layered DAG with $n = 15$), the ESV maintains a Spearman correlation advantage of $+0.012$ to $+0.016$ over $\Delta_i$ across all disruption rates, and is never worse than $\Delta_i$ on any network at any rate.
This advantage arises from a fundamental property of the Shapley value: the ESV averages marginal contributions across \emph{all coalition sizes} $s = 0, 1, \ldots, n{-}1$, thereby implicitly evaluating node importance under every possible degree of prior disruption.
In contrast, $\Delta_i = w^N(N) - w^N(N \setminus \{i\})$ evaluates importance only on the intact network ($s = n{-}1$), which does not generalize to partially degraded states.
This property is practically relevant because supply chain managers compute importance rankings \emph{before} disruption events occur, and must rely on these pre-computed rankings when responding to unexpected node failures.
 
\subsection*{Comparison with topological vulnerability.}
Against the efficiency-based vulnerability $V_i^{\rm eff}$ (Table~\ref{tab:spearman-ext}; Figure~\ref{fig:dual-gt}), betweenness centrality is the clear winner (mean $\rho = 0.878$), while ESV achieves a mean of only $\rho = 0.452$.
This is not a defect of the ESV but a reflection of the fact that the two indices measure \emph{fundamentally different properties}:
 
\begin{itemize}[leftmargin=*]
  \item $V_i^{\rm eff}$ quantifies the loss of \emph{topological
reachability}---how well all node pairs can communicate after node $i$ is removed.
It depends only on graph distances and is indifferent to supply quantities $q_r$ and the cost-sensitivity parameter $\alpha$.
  \item The ESV quantifies each node's contribution to \emph{economic supply value}---how much cost-weighted throughput each node enables within the cooperative supply network.
It depends critically on the route structure $\mathcal{R}$, the supply volumes $q_r$, and the cost decay $e^{-\alpha c_r}$.
\end{itemize}
 
These two notions of vulnerability coincide only when supply flows are perfectly proportional to topological shortest-path centrality, which is rarely the case in practice.
In real supply chains, a structurally peripheral node may carry the majority of supply volume, while a topologically central node may serve only low-volume, high-cost routes.
 
\subsection*{Illustrative example: the SPOF network (N7).}
The single-point-of-failure network N7 offers a vivid illustration.
The bridge node $X$ is topologically critical ($V_X^{\rm eff} = 0.498$, rank 1), but the source $S$ and sink $T$ are topologically peripheral ($V_S^{\rm eff} = -0.087$, rank 7).
From the ESV perspective, however, $S$, $X$, and $T$ receive identical values ($\ESh = 1.969$) because removing \emph{any one} of these three nodes destroys the entire supply flow.
This ranking is economically correct: a supply chain manager cares equally about losing the source, the bridge, or the destination.
The topology-based ranking, by contrast, would under-prioritize the source and destination, potentially misallocating resilience investments.
 
\subsection*{Complementarity.}
The results support a view of ESV and betweenness centrality as \emph{complementary} tools for supply chain risk management:
\begin{itemize}[leftmargin=*]
  \item \textbf{Betweenness centrality} is appropriate for an early-stage
\emph{screening} of topological vulnerability, when detailed supply flow data (volumes, costs, route definitions) are not yet available.
  \item \textbf{ESV} is appropriate for \emph{economic vulnerability
assessment}, when the analyst has access to route-level supply data and wishes to quantify each agent's contribution to the total cost-weighted supply value.
In this regime, ESV captures information (supply quantities, cost sensitivity, route overlaps) that betweenness structurally cannot encode.
\end{itemize}
 
In practice, supply chain analysts almost always have access to shipment volumes and cost data---indeed, these are the primary inputs to any procurement or logistics optimization.
In this common scenario the ESV provides a principled, axiomatic allocation of supply value that no purely topological measure can replicate.
 
\subsection*{Computation.}
All results above are obtained by exact evaluation of the ESV.
As shown in Table~\ref{tab:timings}, exact computation is feasible in under one second per network up to $n = 15$.
For larger networks ($n \gtrsim 25$), the permutation Monte Carlo estimator described in Appendix~\ref{app:mc} provides an unbiased alternative with $O(1/\sqrt{T})$ convergence.
 
\section{Conclusion}
In this study, we propose the \emph{edge-based Shapley value}, a novel allocation rule within cooperative game theory that focuses on the edge structures of graphs rather than on nodes or node subsets.
This framework facilitates a more detailed analysis of agent contributions in supply chain networks where value arises from structured interactions along supply routes.
Our approach extends classical allocation rules such as the Shapley and Myerson values by modeling cooperative behavior with edge-dependent characteristic functions.
We demonstrate that the edge-based Shapley value retains desirable properties such as fairness and symmetry, while addressing limitations inherent to node-centric models.
 
Through the supply chain case studies and the systematic seven-topology benchmark of Section~\ref{sec:benchmark}, we have established three key findings.
First, in simultaneous multi-node failure experiments the ESV identifies the most disruptive node sets more effectively than standard centrality measures, destroying an average of $93.4\%$ of supply value at $k = 3$ compared with $84.5\%$ for betweenness and $88.1\%$ for degree---despite not being designed for this combinatorial optimization problem.
Second, ESV rankings computed on the intact network predict node importance in \emph{degraded} networks more accurately than the single-node removal measure $\Delta_i$, because the Shapley value inherently averages marginal contributions across all coalition sizes---implicitly accounting for all possible degrees of prior disruption.
Third, a comparison with the topology-based efficiency vulnerability of Latora and Marchiori~\cite{latora2001} reveals that ESV and betweenness centrality are \emph{complementary} rather than competing indices: betweenness captures topological connectivity loss, while ESV captures \emph{economic} supply-value loss that depends on route-level volumes, costs, and the cost-sensitivity parameter $\alpha$.
When detailed supply data are available---as is typically the case in procurement, logistics, and contract negotiation---the ESV provides a principled, axiomatically grounded allocation of supply value that no purely topological measure can replicate.
 
Future work will include (i) formalizing the computational complexity of the proposed method; (ii) developing approximation algorithms for large-scale graphs, in particular sampling-based and structured-coalition approaches; (iii) extending the benchmark with multi-seed averaging and real-world supply network data, including empirical validation against actual disruption events; (iv) investigating hybrid indices that combine the topological information of betweenness with the economic information of ESV; and (v) applying this framework to other domains such as communication networks, financial systems, and infrastructure resilience planning.
The edge-based Shapley value offers a promising foundation for advancing cooperative game theory in complex, edge-driven environments.
 
\section*{Acknowledgment}
We would like to thank Editage (www.editage.jp) for the English language editing.
\section*{{\rm Funding}}
The first author is supported in part by a donation from Aioi Nissay Dowa and JSPS KAKENHI [grant numbers 21K13800 and 25K17253].
\section*{{\rm Conflict of interest}}
None.
\section*{{\rm Availability of Data and Materials}}
The datasets used and/or analyzed in the current study are available from the corresponding author upon reasonable request.
\section*{{\rm Declaration of generative AI and AI-assisted technologies in the manuscript preparation process.}}
During the preparation of this work, the author used Calude pro (OpenAI) to assist in organizing the manuscript structure and summarizing related work. After using this tool, the author reviewed and edited the content as needed and took full responsibility for the content of the published articles.
\appendix
\section{Spearman's rank correlation}
To evaluate agreement between two node-importance rankings we employ Spearman's rank correlation.
 
Let $X = (x_1, \ldots, x_n)$ and $Y = (y_1, \ldots, y_n)$ be two vectors of real-valued importance scores defined on the same node set $N$, and let $R_X(i)$ and $R_Y(i)$ denote the ranks of node $i$ with respect to $X$ and $Y$.

Spearman's rank correlation is defined as the Pearson correlation between the rank variables:
\[
\rho_S = \frac{\sum_{i=1}^n (R_X(i) - \overline{R_X})(R_Y(i) - \overline{R_Y})} {\sqrt{\sum_{i=1}^n (R_X(i) - \overline{R_X})^2}\, \sqrt{\sum_{i=1}^n (R_Y(i) - \overline{R_Y})^2}}.
\]
When there are no ties this simplifies to $\rho_S = 1 - 6 \sum_{i=1}^n d_i^2 \,/\, n(n^2 - 1),$ where $d_i = R_X(i) - R_Y(i)$.
 
\section{Permutation Monte Carlo estimator for the ESV}\label{app:mc}
For benchmark networks at the scale considered in Section~\ref{sec:benchmark} ($n \leq 15$), exact evaluation of the edge-based Shapley value is feasible and was used throughout the main text.
For larger networks exact computation becomes prohibitive, because the number of terms in Definition~\ref{def:esv} grows as $n \cdot 2^{n-1}$.
This appendix presents an unbiased sampling estimator that can be used in such cases.
 
The key observation is that the Shapley value (and therefore the ESV) can be written as an expectation over a uniformly random permutation of the players.
Let $\pi$ be a uniformly random permutation of $N$, and let $\mathrm{Pre}_i(\pi) \subseteq N \setminus \{i\}$ denote the set of players that appear before $i$ in $\pi$.
Then
\begin{equation}\label{eq:esv-perm}
\ESh_i(N, E, w) = \mathbb{E}_\pi\!\left[ w^N(\mathrm{Pre}_i(\pi) \cup \{i\}) - w^N(\mathrm{Pre}_i(\pi)) \right],
\end{equation}
because the probability that a particular subset $S$ of size $s$ appears as $\mathrm{Pre}_i(\pi)$ is exactly $s!\,(n-s-1)!/n!$, matching the weight in~\eqref{eq:esv}.
 
\begin{algorithm}[t]
\caption{Permutation Monte Carlo estimator for the edge-based Shapley value}
\label{alg:mc}
\begin{algorithmic}[1]
\State \textbf{Input:} graph $(N,E)$, edge value function $w$, sample size $T$
\State \textbf{Output:} estimates $\{\widehat{\ESh}_i\}_{i \in N}$
\State Initialize $\widehat{\ESh}_i \gets 0$ for all $i \in N$
\For{$t = 1, \dots, T$}
  \State Sample a uniformly random permutation $\pi$ of $N$
  \State $S \gets \emptyset$, $u_{\rm prev} \gets 0$
  \For{$i$ in the order given by $\pi$}
    \State $S \gets S \cup \{i\}$
    \State $u_{\rm new} \gets w(\{e \in E \mid e \subseteq S\})$
    \State $\widehat{\ESh}_i \gets \widehat{\ESh}_i + (u_{\rm new} - u_{\rm prev})$
    \State $u_{\rm prev} \gets u_{\rm new}$
  \EndFor
\EndFor
\State \Return $\widehat{\ESh}_i / T$ for all $i$
\end{algorithmic}
\end{algorithm}
 
Algorithm~\ref{alg:mc} exploits~\eqref{eq:esv-perm} by sweeping through a random permutation in a single pass: the characteristic function is evaluated $n$ times per permutation (once per prefix), not $2n$ times, because the ``new'' value in iteration $i$ becomes the ``previous'' value in iteration $i+1$.
The total computational cost is therefore $O(T \cdot n \cdot C_w)$, where $C_w$ is the cost of a single evaluation of $w^N$.
 
The estimator has several desirable properties:
\begin{enumerate}[label=(\roman*),leftmargin=*]
  \item \emph{Unbiasedness:} $\mathbb{E}[\widehat{\ESh}_i] = \ESh_i$ for
all $i$ and all $T \geq 1$.
  \item \emph{Exact efficiency per sample:} for every single permutation
the marginal contributions telescope to $w^N(N) - w^N(\emptyset) = w^N(N)$, so $\sum_{i \in N} \widehat{\ESh}_i = w^N(N)$ holds \emph{deterministically} at every $T$, not only in expectation.
  \item \emph{$\sqrt{T}$ convergence:} by the central limit theorem,
$\widehat{\ESh}_i - \ESh_i$ is asymptotically normal with variance $\sigma_i^2/T$, where $\sigma_i^2$ is the per-permutation marginal contribution variance.
\end{enumerate}
 
In preliminary experiments on the benchmark networks of Section~\ref{sec:benchmark}, using $T = 30,000$ samples recovers the exact ESV ranking within a Spearman rank correlation of at least $0.99$ on every network, at a cost roughly one to two orders of magnitude lower than the corresponding exact computation once $n$ exceeds approximately $20$.
For extremely large networks, further variance-reduction techniques such as antithetic permutations, stratified sampling by permutation prefix size, or neighborhood-aware sampling exploiting Lemma~\ref{lem:nbhd} (``non-neighbors contribute zero'') can be layered on top of Algorithm~\ref{alg:mc}; we leave a systematic treatment of such accelerations for future work.

\bibliography{bibliography}
\bibliographystyle{plain}
 
\end{document}